\documentclass[runningheads]{llncs}
\usepackage{graphicx}

\usepackage{color,soul,booktabs,eurosym, hyperref}
\usepackage{xurl} 

\begin{document}
    \title{Tell me, what are you most afraid of?\\Exploring the Effects of Agent Representation on Information Disclosure in Human-Chatbot Interaction}
\titlerunning{Information Disclosure in Human-Chatbot Interaction}
%


%

\author{Anna Stock \and
Stephan Schl\"{o}gl\orcidID{0000-0001-7469-4381} \and
Aleksander Groth\orcidID{0000-0003-4698-2321}} 

\authorrunning{Stock et al.}
%
\institute{MCI | The Entrepreneurial School\\Innsbruck, Austria\\
Dept. Management, Communication \& IT \\
\email{stephan.schloegl@mci.edu}\\
\url{https://www.mci.edu}} 

%
\maketitle              
\begin{abstract}
Self-disclosure counts as a key factor influencing successful health treatment, particularly when it comes to building a functioning patient-therapist-connection. To this end, the use of chatbots may be considered a promising puzzle piece that helps foster respective information provision. Several studies have shown that people disclose more information when they are interacting with a chatbot than when they are interacting with another human being. If and how the chatbot is embodied, however, seems to play an important role influencing the extent to which information is disclosed. Here, research shows that people disclose less if the chatbot is embodied with a human avatar in comparison to a chatbot without embodiment. Still, there is only little information available as to whether it is the embodiment with a human face that inhibits disclosure, or whether any type of face will reduce the amount of shared information. The study presented in this paper thus aims to investigate how the type of chatbot embodiment influences self-disclosure in human-chatbot-interaction. We conducted a quasi-experimental study in which $n=178$ participants were asked to interact with one of three  settings of a chatbot app. In each setting, the humanness of the chatbot embodiment was different (i.e., human vs. robot vs. disembodied). A subsequent discourse analysis explored difference in the breadth and depth of self-disclosure. Results show that non-human embodiment seems to have little effect on self-disclosure. Yet, our data also shows, that, contradicting to previous work, human embodiment may have a positive effect on the breadth and depth of self-disclosure.
%
\keywords{Health Chatbots \and Agent Visualization \and Anthropomorphism \and Information Disclosure \and Self-Disclosure.}
\end{abstract}
\section{Introduction and Related Work}\label{sec:intro}
The increasing propagation of voice based conversational assistants such as Apple's \emph{Siri}, Google's \emph{Assistant}, Microsoft's \emph{Cortana} and Amazon's \emph{Alexa}, has also led to an increase in chatbot use. Ever since in 2016 Facebook decided to allow for artificial bots to be integrated into its messaging platform, the number of available text-based chatbots has been growing significantly~\cite{adamopoulou2020chatbots}, so that in 2018, already more than 300,000 chatbots were actively interacting with users via Facebook Messenger\footnote{Online: \url{https://venturebeat.com/2018/05/01/facebook-messenger-passes-300000-bots/} [accessed: February 10\textsuperscript{th} 2023]}. 

\subsection{Chatbot Visual Appearance and Classification}
Chatbots can, but do not have to have a visual appearance (often called agent or avatar representations\footnote{Note: strictly speaking the visual representation of a chatbot should be called `agent' since it is controlled by an algorithm whereas a human-controlled visual appearance should be referred to as `avatar'~\cite{fox2015avatars}}). Visualization options range from static and interactive virtual to physically completely embodied agents\cite{diederich2019conversational}. These differences in presentation also offer the possibility to integrate various non-verbal features, which may influence how users perceive chatbots~\cite{appel2012does} and consequently interact with them.

Beyond their appearance, chatbots can be classified in different ways. Work by Adamopoulou and Moussiades~\cite{adamopoulou2020chatbots}, based on Nimavat \& Champaneria~\cite{nimavat2017chatbots}, for example, proposes a classification according to seven key aspects. These include: (1) the subject area in which a chatbot must be able to `converse' (i.e., overarching or topic-specific); (2) the type of service a chatbot provides (i.e., offer company, provide a specific service, or allow for a connection to another human and/or chatbot); (3) the goal of the interaction (provide information, have a conversation, fulfil a task); (4) the way the answers have to be generated; (5) the level of required human intervention; (6) the licensing (open source vs. commercial); and (7) the communication channel (text, language, images or a combination thereof). 

\subsection{Chatbots in Health Care and Therapy}
While there is a general increase in chatbot use, it is particularly the health domain that has seen strong uptakes throughout recent years. Particularly with respect to mental health, experts put great hope in conversational technology~\cite{d2020ai}. In fact, chatbots have already shown promising results in several studies both clinical~\cite{kang2010virtual,oh2020efficacy} and non-clinical~\cite{fitzpatrick2017delivering}. Measurable success was particularly evident in the support of people in relation to psycho-education and adherence. That is, chatbots are deemed effective in helping people help themselves by offering assistance to access relevant information and therapy material~\cite{vaidyam2019chatbots,gardiner2017engaging}. While health apps can help with meditation (e.g., \emph{Headspace}\footnote{Online: \url{https://www.headspace.com/} [accessed: February 10\textsuperscript{th} 2023]}, \emph{Calm}\footnote{Online: \url{https://www.calm.com/} [accessed: February 10\textsuperscript{th} 2023]}), recovery from addiction (e.g., \emph{Recovery Record}\footnote{Online: \url{https://www.recoveryrecord.eu/} [accessed: February 10\textsuperscript{th} 2023]}) or depression (e.g, \emph{Talkspace}\footnote{Online: \url{https://www.betterhelp.com/} [accessed: February 10\textsuperscript{th} 2023]}), the more advanced chatbot systems may further engage in simple conversations and thus even offer basic therapy~\cite{d2020ai}.  

Although chatbot conversations may not replace therapy sessions, proponents argue that they can provide support and have a supplementary and relieving effect on mental health~\cite{d2020ai}. Following the rules of cognitive behavioral therapy, successful examples include \emph{Woebot}\footnote{Online: \url{https://woebothealth.com/} [accessed: February 10\textsuperscript{th} 2023]}~\cite{monnier2020woebot}, \emph{Wysa}\footnote{Online: \url{https://www.wysa.io/} [accessed: February 10\textsuperscript{th} 2023]}~\cite{inkster2018empathy} or \emph{Tess}\footnote{Online: \url{https://www.x2ai.com/individuals} [accessed: February 10\textsuperscript{th} 2023]}. It is particularly the `around-the-clock availability' and the low entry threshold of these conversational tools that helps to reach people who are often reluctant to seek help from therapy –- whether due to cost or due to social stigma~\cite{LUCAS201494,vaidyam2019chatbots}. 

\subsection{Information Disclosure}
To clarify treatment of psychological and physical illnesses, disclosure plays a major role. Griffin describes disclosure as \textit{``the voluntary sharing of personal history, preferences, attitudes, feelings, values, secrets, etc., with another person''}~\cite[p. 114]{griffin2003first}, whereas Cozby summarizes it as \textit{``any information about oneself that a person A verbally communicates to a person B''}~\cite[p. 73]{cozby1973self}. Finally, Pickard \& Roster refer to it as \textit{``the truthful, sincere, and intentional communication of private information about oneself or others that makes oneself or others more vulnerable''}~\cite[p. 2]{pickard2020using}.

Social Penetration Theory (SPT) describes the development of interpersonal relationships, where an initially superficial relationship becomes increasingly intimate~\cite{altman1973social}. In SPT, disclosure is described as consisting of two dimensions, i.e. breadth and depth of disclosure. Depth describes the degree of intimacy in a relationship and the degree to which a particular area of life is discussed. Breadth describes the range of areas of a person's life that are talked about. To this end, it is possible for a person to be very open in one area, while not talking about another area. The higher the degree of intimacy, the more areas can be disclosed at the same time~\cite{griffin2003first}.

Disclosure usually happens when the expected benefits outweigh the expected risks~\cite{doi:10.1080/03637750902828412}.Here Omarzu's Disclosure Decision Model~\cite{omarzu2000disclosure} describes the path to disclosure as a multi-step decision process: (1) evaluate whether a social benefit may be achieved (e.g., social affirmation, intimacy, relief in case of distress or despair, social control, etc.); (2) evaluate whether disclosure is a plausible means of achieving this goal; and (3) assess the potential risk of disclosure (e.g., potential rejection, weakening of one's own autonomy and personal integrity, embarrassment, etc.). 

In health contexts, the challenge is usually to achieve the highest possible degree of disclosure. The underlying motivation may be to obtain important information~\cite{LUCAS201494} or to benefit from the positive effects that disclosure produces in interactions. Ho and colleagues summarize these positive effects found in previous studies into three categories: the improvement of short-term emotional experience (stress, anxiety, reduction), which can ultimately lead to long-term mental health improvement, the improvement of relationship quality through the promotion of closeness and intimacy, and the improvement of the psychological condition by promoting self-esteem and affirmation~\cite{ho2018psychological}. 

Building on previous research in decision making related to information disclosure Greene developed the Disclosure Decision-making Model, which focuses on disclosure in the health context -- in this case, particularly on how people deal with the disclosure of information after a disease diagnosis or in the course of medical treatment~\cite{greene2015integrated}. Empirical analyses found that the efficacy of information disclose is particularly important. If an individual assumes that the response to the disclosure is helpful and that the disclosure therefore effective, the willingness to disclose increases significantly~\cite{greene2012assessing}. 

Several studies comparing different conversational settings furthermore found, that people are willing to disclose more about themselves when interacting with a chatbot or other form of conversational agent than when talking to a human~\cite{lind2013survey,LUCAS201494,PICKARD201623}. To this end, chatbots may be considered media agents which are capable of social interaction~\cite{gambino2020building}. 

\subsection{Chatbots as Social Actors}
According to the CASA-Framework (i.e., \textit{Computers Are Social Actors}) and its underlying media equation, chatbots should trigger natural responses similar to those triggered by human interlocutors~\cite{nass1997machines}. Here anthropomorphism may help in the establishment and maintenance of human-chatbot relationships~\cite{bickmore2005establishing}, enhance the perceived social presence of the chatbot~\cite{ARAUJO2018183}, and have a positive effect on customer satisfaction~\cite{verhagen2014virtual}. Of particular importance in this regard are the form in which the chatbot is presented (i.e., the visual cues) and the degree of human resemblance with which this is done. That is, while anthropomorphism in the representation favors trust in chatbots~\cite{de2016almost} and increases its social presence~\cite{sah2015effects}, people seem to disclose more to chatbots when they are represented in a less anthropomorphic way~\cite{bailenson2006effect}. The level of disclosure typically decreases when a chatbot with a human face is presented~\cite{lind2013survey,pickard2020using}. This suggests that people feel less social risk when disclosing information to a non-human like entity. 

Another negative effect of anthropomorphism is shown in the so-called `Uncanny Valley' effect~\cite{mori2012uncanny}, which describes the reaction of humans to anthropomorphic entities. Accordingly, if an entity is almost, but not quite human-/lifelike, the human response to it abruptly changes from empathy to rejection. Despite these negative effects of human-like appearance, research has also shown that faces may encourage disclosure as they foster social connection~\cite{knapp2013nonverbal} -- even, if they are not human~\cite{Rosenthal-vonderPutten2013}. 

Considering both positive and negative effects of anthropomorphic visual cues, Pickard \& Roster~\cite{pickard2020using} thus argue that disclosure to a chatbot may be highest when it is presented with a face that humans can connect with, but at the same time does not have such a strong human resemblance that it would risk causing feelings of eeriness. Our work aimed to investigate this assumption.

\section{Theoretical Framework and Hypothesis Development}\label{sec:relwork}
Following the definition by Gambino et al.~\cite{gambino2020building} we may assume for chatbots to fulfill the criteria to be considered media agents. Consequently, we may further assume that humans show similar social reactions towards chatbots as they do towards humans.

Various studies have furthermore shown that anthropomorphism, i.e., the degree to which a chatbot is perceived to exhibit human characteristics, has a significant influence on the strength of certain reactions. Anthropomorphism can, for example, promote the establishment and maintenance of long-term human-chatbot relationships~\cite{bickmore2005establishing}, increase the perceived social presence of a chatbot~\cite{ARAUJO2018183}, and have a positive effect on customer satisfaction~\cite{verhagen2014virtual}.

The chatbot's representation (i.e., its visual cues) and its degree of human resemblance are of particular importance here. That is, while anthropomorphism in the representation favors trust in a chatbot~\cite{de2016almost} and increases its social presence~\cite{sah2015effects}, the effect in other areas can be be negative. For example, people disclose more to a chatbot when it is presented in a less realistic way~\cite{bailenson2006effect}. And in general, the level of disclosure typically decreases when a chatbot is presented with a human face~\cite{lind2013survey,pickard2020using}. This suggests that people would feel to be subject to the same social risk when conversing with a chatbot as they feel when disclosing in a conversation with a human, especially if the chatbot is presented with a human-like face. 

On the other hand, the visual representation of a face can also encourage disclosure. That is, people can build a social connection to a face~\cite{knapp2013nonverbal}, even if it is not human~\cite{Rosenthal-vonderPutten2013}. This social connection in turn encourages disclosure~\cite{altman1973social}, unless its representation taps into Mori's `Uncanny Valey'~\cite{mori1970uncanny}. Here, an imperfect almost human-like representation starts to trigger feelings of eeriness~\cite{macdorman2006uncanny}. Several studies have shown that the uncanny valley effect also occurs in human-chatbot interaction (cf.,\cite{ciechanowski2019shades,macdorman2009too,thaler2020agent}). 

Considering the positive effect of faces on the one hand and the potential negative effect of the uncanny valley on the other hand, Pickard \& Roster~\cite{pickard2020using} thus argue that a chatbot may work most effective when it offers a face that is not too human-like. From this, we derive the following hypothesis:

\begin{center}
    \textit{H1: The degree of people's information disclosure is higher when questions are asked by a ``non-embodied'' (i.e. not visually represented) chatbot or a chatbot that has only some human-like appearance than by a chatbot whose representation shows a high degree of human resemblance.}
\end{center}

In order to test this hypothesis we designed a study in which participants were asked to interact with one of three chatbots and answer a number of questions. 

\section{Methodology}\label{sec:methodology}
\begin{figure}[ht]
    \centering
    \includegraphics[width=0.8\textwidth]{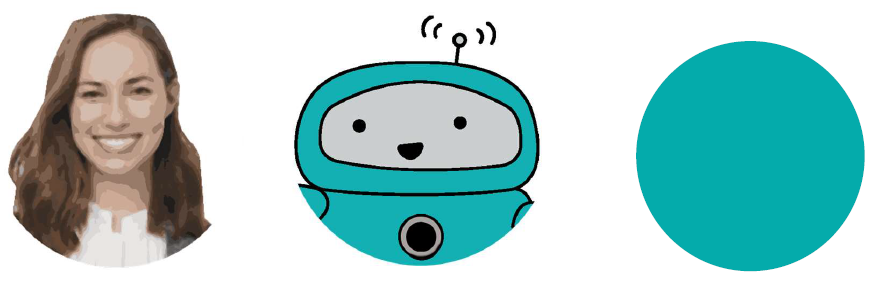}
    \caption{Chatbot designs to investigate the effect of visual representation on information disclosure. The left image shows high human resemblance, the image in the middles low human resemblance, and the image on the right no form of embodimnent}
    \label{fig:chatbots}
\end{figure}

Our goal was to evaluate the effect the visual representation of chatbots has on people's information disclosure behavior. To do so, we used a survey-setup in which participants were asked to interact with one of three chatbot designs in order to answer personal questions. We then evaluated people's responses in light of the chatbot design they were interacting with. The following three designs were used (cf. Figure~\ref{fig:chatbots}):

\begin{enumerate}
    \item High human resemblance based on a slightly stylized photo of a woman (i.e., human-like)
    \item Low human resemblance depicted by a robot pictorial (i.e., robot-like)
    \item Without any form of `embodiment' (i.e, none-embodied)
\end{enumerate}

Chatbots were available online and optimized to be accessible through either PC or mobile devices. The language and expression as well as the typeface which was used to display the dialog were standardized and the same for all the three chatbot designs. Slightly delayed response times aimed to increase the perceived social presence of the chatbots~\cite{gnewuch2018faster}. We used both open and closed questions, all of which required some form of personal disclosure. Closed questions were inspired by Lind et al.~\cite{lind2013survey} and it was expected that people would be honest when reporting on negative things about themselves~\cite{kreuter2008social}. Open questions were taken from Pickard \& Roster~\cite{pickard2020using}, who also investigated depth and breadth of information disclosure in different interview settings. Simiarly, we used the `Objective Information Disclosure' (OID) scale to measure the depth and the number of words~\cite{joinson2001knowing,kang2010virtual} to evaluate the breath of the disclosure in these questions. 

With respect to the type of information that was requested, questions were classified as \textit{demographic}, \textit{perceptive} and \textit{informative} (cf. Section~\ref{li:questions}). Demographic questions were collecting demographic information and used to investigate group differences between participants' answers to the other questions. Perceptive questions focused on the perceived level of human-likeness attached to our three chatbot designs, and informative questions explored participants' information disclosure behaviour. All questions were reviewed and approved by the school's Research Ethics Committee in terms of ethical considerations regarding research with human participation. Also, none of the questions were compulsory so that participants could decide for themselves whether they felt comfortable providing an answer or not.

\section{Instruments}
\subsection{Questions}
The following questions were asked by the chatbot during the interaction with a participant:

\begin{itemize}\label{li:questions}
    \item Demographic Questions:
    \begin{enumerate}
        \item \textit{How old are you?} [\textless 18 $\vert$ 18--35 $\vert$ \textgreater 35]
        \item \textit{Where do you live?} [Austria $\vert$ Germany $\vert$ Italy $\vert$ Switzerland $\vert$ Other]
        \item \textit{What is your occupation?} [Student $\vert$ Employed/Self-employed $\vert$ Unemployed $\vert$ Homemaker $\vert$ Other]
        \item \textit{What is your gender?} [Male $\vert$ Female $\vert$ Diverse $\vert$ Do not want to say]
    \end{enumerate}
    \vspace{0.5cm}
    \item Perceptive Questions:
    \begin{enumerate}
        \item \textit{The picture looks human-like:} [I totally disagree $\vert$ I disagree $\vert$ neutral $\vert$ I agree $\vert$ I totally agree]
        \item \textit{The picture looks realistic:} [I totally disagree $\vert$ I disagree $\vert$ neutral $\vert$ I agree $\vert$ I totally agree]
        \item \textit{The picture looks cartoon-like:} [I totally disagree $\vert$ I disagree $\vert$ neutral $\vert$ I agree $\vert$ I totally agree]
    \end{enumerate}
    \vspace{0.5cm}
    \item Informative Questions:
    \begin{enumerate}
        \item \textit{How often do you read a newspaper (both printed and online)?} [never $\vert$ less than once a week $\vert$ once a week $\vert$ several times during the week $\vert$ daily]
        \item \textit{What is it that you like to do with your closest friends or family?} [\dots]
        \item \textit{Which of the following describes your money saving habits best:} [I don't save money -- I usually spend more than I have $\vert$ I don't save money -- I usually spend as much as I have $\vert$ I save what is left at the end of the month -- no regular saving plan $\vert$ We save the income of one family member -- the rest we spend $\vert$ I spend my regular income and save additional, irregular income $\vert$ I save regularly every month]
        \item \textit{What is it that you dislike the most about your look?} [\dots]
        \item \textit{How often during the last 12 months have you in a bus or public place offered your seat to a stranger?} [never during the last 12 months $\vert$ once during the last 12 months $\vert$ two to three times during the last 12 months $\vert$ once per month $\vert$ once per week $\vert$ more than once per week $\vert$ I don't know]
        \item \textit{How often during the last 12 months have you given money to a homeless person?} [never during the last 12 months $\vert$ once during the last 12 months $\vert$ two to three times during the last 12 months $\vert$ once per month $\vert$ once per week $\vert$ more than once per week $\vert$ I don't know]
        \item \textit{What are you most scared of?} [\dots]
        \item \textit{Think about a person that you love. Why do you love this person?} [\dots]
    \end{enumerate}
\end{itemize}

\subsection{OID Analysis Scheme}
The following evaluation scheme was used to analyze participants' responses to the chatbot: \textit{The response to the question contains\dots} [1 = Not at all, 5 = A great deal]:

\begin{itemize}\label{li:oid}
    \item \textit{shame.}
    \item \textit{concrete details.}
    \item \textit{an answer to the question asked.}
    \item \textit{embarrassing information.}
    \item \textit{sensitive information.}
    \item \textit{information that violates social norms.}
    \item \textit{information that increases the person’s risk for harm or negative consequences.}
    \item \textit{an admission of the person’s imperfection.}
    \item \textit{an admission of guilt.}
    \item \textit{an admission of failure on the person’s part.}
    \item \textit{negative emotions.}
    \item \textit{positive emotions.}
\end{itemize}

\section{Results}
The link to the chatbot-driven survey was distributed via social networks, mainly focusing on people under the age of 35 living in the DACH region (i.e. Germany, Austria, Switzerland, North of Italy). Data from respondents who were under the age of 18 was excluded from the analysis and consequently deleted straight away. Answers from $n=178$ people aged 18 or above were collected and evaluated ($113$ female, $62$ male, $1$ diverse, $2$ did not want to say). Of these, 158 were in the 18-35 age group and 20 were over 35 years old. Participants were randomly assigned to one of the three chatbot designs. This yielded data from $60$ interactions with the human-like chatbot (human), $66$ interactions with the robot-like chatbot (robot) and $52$ interactions with the non-embodied chatbot (no embodiment).

Similar to Nowak \& Rauh~\cite{nowak2005influence}, we used three question items to evaluate the human-like appearance of the chabtbots (cf., Section~\ref{li:questions}). As expected, the data shows that the photo of the stylized women was perceived as being significantly more anthropomorphic ($M=3.25, SD=0.86$) than the robot pictorial ($M=1.40, SD=0.60$); $t(177)=–26.03, p<0.001, r=0.890$. 

Investigating differences in information disclosure, the data shows that participants state to have more often offered a seat to a stranger (informative question 5; cf. Section~\ref{li:questions}) when the question was asked by the human-like chatbot representation than by the one without embodiment; $Mann–Whitney \: U=315, p=0.006, r=0.341$. Further exploring the depth of participant answers we used Pickard \& Roster's~\cite{pickard2020using} answer analysis scheme to calculate an average OID scale (cf. Section~\ref{li:oid}) for each of of the open questions (i.e., informative question 2, 4, 7 and 8; cf. Section~\ref{li:questions}). The data shows more depth in answers to the human-like chatbot compared to the robot-like chatbot with respect to the questions about people's greatest fear (informative question 7; cf. Section~\ref{li:questions}); $t(118)=3.17, p=0.002, r=0.280$. A comparison between the human-like chatbot and the none-embodied chatbot showed a similar although less significant difference; $t(103)=2.14, p=0.035, r=0.206$. Other than this, the data does not point to any differences concerning the depth of participants' answers. 

Looking at the breath of provided answers we compared the number of words participants wrote. Also here, it was the question about people's greatest fear (informative question 7; cf. Section~\ref{li:questions}) where the data shows a significant difference in answering behaviour. That is, the human-like chatbot representation triggered more words in participants' answers than the robot-like representation ($t_{Welsch}(78.38)=2.31, p=0.024, r=0.252$) or the none-embodied representation ($t_{Welsch}(93.56)=2.15, p=0.034, r=0.217$). 

Looking at the answer behavior, it can further be seen that the robot-like chatbot representation produced the highest number of missing or elusive answers to questions. That is, $15.15\%$ of participants interacting with the robot-like chatbot left at least one question empty or did not really answer it, compared to $11.54\%$ of participants interacting with the none-embodied chatbot and $10.00\%$ of participants interacting with the human-like chatbot. 


\section{Discussion of Results and their Limitations}\label{sec:results}
Our investigation evaluated the extent to which the human resemblance of chatbots influences information disclosure. The collected data was subjected to a discourse analysis. It was expected that disclosure would be highest with little human resemblance, compared to high human resemblance and none-embodiment. Analysis results, however, suggest the contrary. That is, the analysis on breadth and depth of disclosure indicates that disclosure to a chatbot with high human resemblance is higher than to a chatbot with less human resemblance or without embodiment. The latter, in particular, contradicts previous work by Pickard \& Roster~\cite{pickard2020using}, Joinson~\cite{joinson2001knowing}, and Lind et al.~\cite{lind2013survey}. Reasons for this may be found in our research design. Most studies on disclosure in chatbot settings work with monitored experiments in which participants communicate with a system via spoken language. Our study, however, focused on text-based interaction.

A study by Hill et al. for example found that people communicate with chatbots in written form differently than they communicate in spoken language~\cite{hill2015real}. At the same time, the setting, in which the respondents answered the questions (i.e., in private and without supervision) may have had an effect. For example, a study by Li et al.~\cite{li2019self} found that people in the private sphere are more willing to disclose. It is therefore possible that our participants were generally more open to disclosure and that accordingly fewer differences triggered by chatbot representations were found.

Furthermore, in contrast to other studies on the subject (e.g.,~\cite{kang2010virtual,lind2013survey,lucas2017reporting,pickard2020using,astrid2010doesn}), the human-like representation in our study was static. The lack of animation may have alleviated a potential uncanny valley effect and increased the acceptance of the more anthropomorphic representation.

Finally, a more frequent use of emoticons with our embodied (i.e., human-like and robot-like) chatbots does suggest that both representations received a higher degree of liking than the non-embodiment representation. Altman \& Taylor's~\cite{altman1973social} social penetration theory assumes that relationships become deeper and more intimate over time and that disclosure occurs after a personal `cost-benefit analysis'. Our study was not designed to produce any benefit for participants so one may argue that they had little reasons to disclose information. A long-term study design thus may have produced a more diverse answering behaviour. 

\section{Conclusion and Future Outlook}\label{sec:conclusion}
We have reported on the results of a chatbot-driven survey which investigated people's information disclosure behaviour with respect to different types of chatbot representation. Results have shown that, contrary to previous work, our human-like representation was not affected by the uncanny valley and able to trigger responses with a higher level of breath and depth -- at least when it come to people reporting on their greatest fear.

These results, however, are subject to a number of limitations.  First, they are based on rather small samples sizes of approx. 60 people per survey setting. Second, we furthermore assumed that the interaction with the different chatbots is significantly influenced by the social reactions their representations trigger in participants -- be it in relation to the uncanny valley effect, sympathy or the establishment of a social connection. Since no data was collected that could provide clear information on what triggered responses, manipulations could have had a different effect. Finally, the evaluation of disclosure via the OID scale assumes that participants adhere to the same social norms and have similar feelings in relation to the disclosure of information. It disregards the fact that people, depending on their life experience or value system, may regard the sensitivity of information differently and therefore choose different levels of disclosure~\cite{pickard2020using}.

To tackle some of these limitations, future work, should use a greater sample size and focus on long-term studies of human-chatbot interaction so as to investigate how social penetration over time effects information disclosure. It furthermore should use post-survey questionnaires to record emotional attitudes of participants towards the the different chatbot representations. And finally, with respect to chatbot representation, we recommend exploring the effect of chatbot gender stereotypes (e.g,~\cite{chaves2021should}) or the influence of various ethnic representations (e.g.,~\cite{ruane2019conversational}).

\bibliography{hcii}
\bibliographystyle{splncs04}
\end{document}